\documentclass[twocolumn, superscriptaddress,aps,floatfix,tightenlines,amssymb]{revtex4}
\usepackage{epsfig}
\usepackage{amsfonts}
\usepackage{amsmath}
\usepackage{epsfig}
\usepackage{comment}

\usepackage{color}

\newcommand{\ket}[1]{\left | #1 \right \rangle}
\newcommand{\bra}[1]{\left \langle #1 \right |}

\begin{document}

\title{Realization of a Knill-Laflamme-Milburn C-NOT gate \\
--a photonic quantum circuit combining effective optical nonlinearities}

\author{Ryo Okamoto}
\affiliation{%
Research Institute for Electronic Science, Hokkaido University,
Sapporo 060-0812, Japan
}%
\affiliation{%
The Institute of Scientific and Industrial Research, Osaka University,
Mihogaoka 8-1, Ibaraki, Osaka 567-0047, Japan
}%

\author{Jeremy L. O'Brien}
\affiliation{Centre for Quantum Photonics, H. H. Wills Physics Laboratory \& Department of Electrical and Electronic Engineering, University of Bristol, Merchant Venturers Building, Woodland Road, Bristol, BS8 1UB, UK
}%

\author{Holger F. Hofmann}
\affiliation{%
Graduate School of Advanced Sciences of Matter, Hiroshima University, Hiroshima 739-8530, Japan
}%

\author{Shigeki Takeuchi}
\altaffiliation[Electronic address: ]{takeuchi@es.hokudai.ac.jp}
\affiliation{%
Research Institute for Electronic Science, Hokkaido University,
Sapporo 060-0812, Japan
}%
\affiliation{%
The Institute of Scientific and Industrial Research, Osaka University,
Mihogaoka 8-1, Ibaraki, Osaka 567-0047, Japan
}%

\maketitle

\noindent\textbf{Quantum information science addresses how uniquely quantum mechanical phenomena such as superposition and entanglement can enhance communication \cite{gi-nphot-1-165}, information processing \cite{la-nat-464-45} and precision measurement \cite{gi-sci-306-1330}. Photons are appealing for their low noise, light-speed transmission and ease of manipulation using conventional optical components \cite{ob-nphot-3-687}. However, the lack of highly efficient optical Kerr nonlinearities at single photon level was a major obstacle. In a breakthrough, Knill, Laflamme and Milburn (KLM) showed that such an efficient nonlinearity can be achieved using only linear optical elements, auxiliary photons, and measurement\cite{kn-nat-409-46}. They proposed a heralded controlled-NOT (CNOT) gate for scalable quantum computation using a photonic quantum circuit to combine two such nonlinear elements. Here we experimentally demonstrate a KLM CNOT gate. We developed a stable architecture to realize the required four-photon network of nested multiple interferometers based on a displaced-Sagnac interferometer and several partially polarizing beamsplitters. This result confirms the first step in the KLM `recipe' for all-optical quantum computation, and should be useful for on-demand entanglement generation and purification. Optical quantum circuits combining giant optical nonlinearities may find wide applications across telecommunications and sensing.}

Several physical systems are being pursued for quantum computing \cite{la-nat-464-45}---promising candidates include trapped ions, neutral atoms, nuclear spins, quantum dots, superconductor and photons---while photons are indispensable for quantum communication \cite{gi-nphot-1-165} and are particularly promising for quantum metrology \cite{na-sci-316-726}. In addition to low-noise quantum systems (typically two-level `qubits') quantum information protocols require a means to interact qubits to generate entanglement. The canonical example is the CNOT gate, which flips the state of the polarisation of the `target' photon conditional on the `control' photon being horizontally polarized (the logical `1' state). The gate is capable of generating maximally entangled two-qubit states, which together with one-qubit rotations provide a universal set of logic gates for quantum computation. 

The low noise properties of single photon qubits are a result of their negligible interaction with the environment, however, the fact that they do not readily interact with one-another is problematic for the realization of a CNOT or other entangling interaction.  Consequently it was widely believed that matter systems, such as an atom or atom-like system \cite{tu-prl-75-4710}, or an ensemble of such systems \cite{sc-oe-21-1936}, would be required to realize such efficient optical nonlinearities. Indeed the first proposals for using linear optics to benchmark quantum algorithms require exponentially large physical resources \cite{Takeuchi1996, ce-pra-57-1477}.

In 2001, KLM made the surprising discovery that a scalable quantum computer could be built from only linear optical networks, and single photon sources and detectors \cite{kn-nat-409-46}. In fact, it was even surprising to KLM themselves, as they had initially intended to proove the opposite. The KLM `recipe' consists of two parts: an optical circuit for a CNOT gate using linear optics, single photon sources \cite{sh-nphot-1-215}, and photon number resolving detectors \cite{ki-apl-74-902}; and a scheme \cite{go-nat-402-390} for increasing the success probability of this CNOT gate ($P=1/16$) arbitrarily close to unity, which harnesses quantum teleportation \cite{be-prl-70-1895} with linear optics \cite{bo-nat-390-575}. This epoch-making result opened the door to the linear optics quantum computation and has spurred a worldwide theoretical and experimental effort to realize such a device \cite{Kok2007}, as well as new quantum communication schemes \cite{gi-nphot-1-165} and optical quantum metrology \cite{na-sci-316-726}.

While a number of quantum logic gates inspired by the KLM approach have been demonstrated \cite{pi-pra-68-032316,ob-nat-426-264,sa-prl-92-017902,ga-prl-93-020504,zh-prl-94-030501,ok-prl-95-210506,ba-prl-98-170502,po-sci-320-646}, none of these gates used the original KLM proposal of a simple measurement induced nonlinearity: either the gates are not heralded (the resultant output photons themselves have to be measured and destroyed) or rely on additional entanglement effects; 
as  we explain below, the KLM scheme is based on a direct implementation of the non-linear sign-shift (NS) gate that relies on the interaction with a single auxiliary photon at a beam splitter. It is thus based on the efficient optical nonlinearity induced by single photon sources and detectors. While a measurement induced nonlinearity has been verified by a conditional phase shift \cite{sa-prl-92-017902}, the technical difficulty of realizing quantum circuits that can combine such elementary quantum operations into a single gate has prevented the implementation of the KLM approach. Specifically, it is a challenging task to implement the nested interferometers needed to perform the multiple classical and quantum interferences that form the elements of the quantum gate operation.



%
The key element in the KLM CNOT gate is the nondeterministic nonlinear sign-shift (NS) gate (Fig. 1a), which operates as follows: When a superposition of the vacuum state $\ket{0}$, one photon state $\ket{1}$ and two photon state $\ket{2}$ is input into the NS gate, the gate flips the sign (or phase) of the probability amplitude of the $\ket{2}$ component: $\ket{\psi} = \alpha\ket{0} + \beta\ket{1} +\gamma\ket{2} \rightarrow \ket{\psi'} = \alpha\ket{0} + \beta\ket{1} -\gamma\ket{2}$. Note that this operation is nondeterministic---it succeeds with probability of $P=1/4$---however, the gate always gives a signal (photon detection) when the operation is successful.
\begin{figure}[t!]
\begin{center}
\includegraphics*[width=6.5cm]{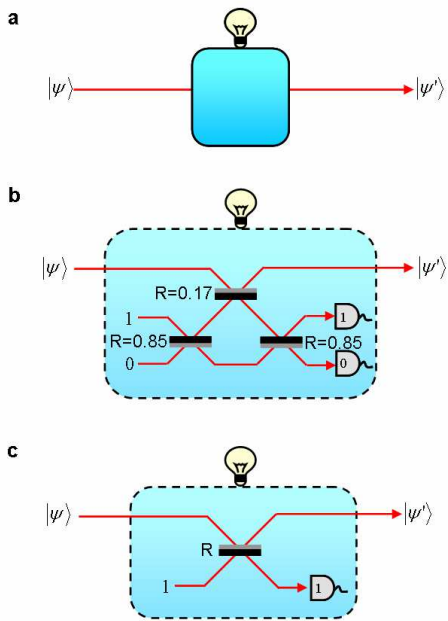}
\vspace{-0.3cm}
\caption{The KLM nonlinear sign-shift (NS) gate. (a) 
If the NS gate succeeds it is heralded; indicated conceptually by the light globe. (b) The original KLM NS gate is heralded by detection of a photon at the upper detector and no photon at the lower detector. Gray indicates the surface of the BS from which a sign change occurs upon reflection.(c) A simplified KLM NS gate for which the heralding signal is detection of one photon.}
\end{center}
\vspace{-0.8cm}
\label{schem1}
\end{figure}

A CNOT gate can be constructed from two NS gates as shown schematically in Fig. 2a \cite{kn-nat-409-46}. Here the control and target qubits are encoded in optical mode or path (`dual-rail encoding'), with a photon in the top mode representing a logical 0 and in the bottom a logical 1. The target modes are combined at a 1/2 reflectivity BS (BS3), interact with the control 1 mode via the central Mach-Zehnder interferometer (MZ), and are combined again at a 1/2 reflectivity BS (BS4) to form another MZ with the two target modes, whose relative phase is balanced such that, in the absence of a control photon, the output state of the target photon is the same as the input state. The goal is to impart a $\pi$ phase shift in the upper path of the target MZ, conditional on the control photon being in the 1 state such that the NOT operation will be implemented on the target qubit. 
When the control input is 1, quantum interference \cite{ho-prl-59-2044} between the control and target photon occurs at BS1:
$\ket{1}_{C_1}\ket{1}_{T_0}\rightarrow\ket{2}_{C_1}\ket{0}_{T_0}-\ket{0}_{C_1}\ket{2}_{T_0}$.
In this case the NS gates each impart a $\pi$ phase shift to these two photon components: $\ket{2}_{C_1/T_0}\rightarrow-\ket{2}_{C_1/T_0}$. At BS2 the reverse quantum interference process occurs, separating the photons into the $C_1$ and $T_0$ modes, while preserving the phase shift that was implemented by the NS gates. In this way the required $\pi$ phase shift is applied to the upper path of the target MZ, and so CNOT operation is realized.



\begin{figure}[t!]
\begin{center}
\includegraphics*[width=7.5cm]{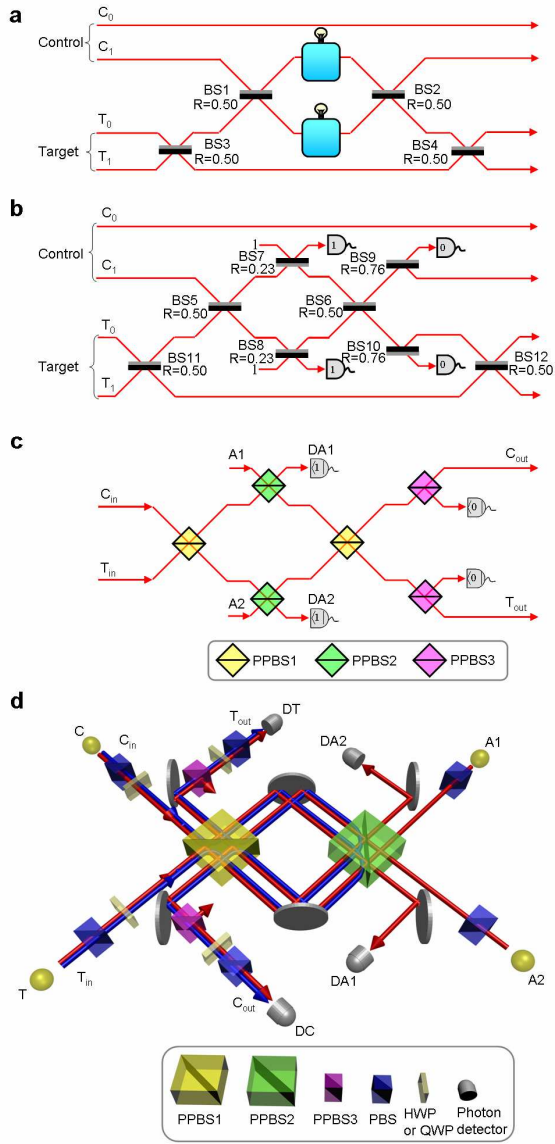}
\vspace{-0.3cm}
\caption{The KLM CNOT gate. (a) The gate is constructed of two NS gates; the output is accepted only if the correct heralding signal is observed for each NS gate. Gray indicates the surface of the BS from which a sign change occurs upon reflection. (b) The KLM CNOT gate with simplified NS gate. (c) The same circuit as (b) but using polarization encoding and PPBSs. (d) The stable optical quantum circuit used here to implement the KLM CNOT gate using PPBSs and a displaced Sagnac architecture. 
The target MZ, formed by BS11 and BS12 in Fig.~\ref{schem2}b, can be conveniently incorporated into the state preparation and measurement, corresponding to a change of basis, as described in the caption to Fig.~\ref{results}.
The blue line indicates optical paths for vertically polarized components, and the red line indicates optical paths for horizontally polarized components.
}
\end{center}
\vspace{-0.8cm}
\label{schem2}
\end{figure}

An NS gate can be realized using an optical circuit consisting of three beam splitters, one auxiliary single photon, and two photon number resolving detectors (Fig. 1b) \cite{kn-nat-409-46}. 
The NS gate is successful, \emph{i.e.} $\ket{\psi}\rightarrow\ket{\psi'}$, when one photon is detected at the upper detector and no photons at the lower detector. 
This outcome occurs with probability 1/4 and so the success probability of the CNOT gate is $(1/4)^2=1/16$.

The key to NS gate operation is multi-photon quantum interference, which can be understood by considering the simplified NS gate shown in Fig. 1c \cite{ra-pra-65-012314}. The probability amplitude for one photon to be detected at the output detector (which is the success signal) can be calculated by summing up the amplitudes of the indistinguishable processes leading to this result: For the $\ket{0}$ input only reflection of the auxiliary photon contributes and the amplitude is simply given by $\sqrt{R}$, where $R$ is the reflectivity of the beamsplitter. For the $\ket{1}$ input the total probability amplitude $1-2R$ is given by the sum of the probability amplitudes for two photons to be reflected ($-R$) and two photons to be transmitted ($1-R$). Finally for the $\ket{2}$ input the probability amplitude is $\sqrt{R}(3R-2)$. 
This shows that nonlinear sign flip of the $\ket{2}$ term, required for NS gate operation, is possible for any $R<2/3$, however, the amplitudes of the $\ket{0}$, $\ket{1}$ and $\ket{2}$ components are also modified by the operation, which is not desired. 
In the original NS gate (Fig. 1b), the path interferometer is used to balance these amplitudes. To preserve these amplitudes in the case where the simplified NS gates are used small losses can be deliberately introduced in the output using BS9 and BS10 in Fig. 2b \cite{ra-pra-65-012314}, at the cost of reducing the success probability slightly (from 0.25 to 0.23), but with the benefit of removing the need for the interferometer in the NS gates. Even with this simplification significant technical difficulties remain: nested interferometers, two auxiliary photons, and several classical and quantum interference conditions. 



We designed the inherently stable architecture shown in Fig. 2d to implement the KLM CNOT gate of Fig. 2b, using polarization to encode photonic qubits. This design takes advantage of two recent photonic quantum circuit techniques: partially polarizing beam splitters \cite{ok-prl-95-210506,ok-sci-323-483} (PPBSs), which results in the circuit shown in Fig. 2c, and the displaced-Sagnac architecture \cite{ok-sci-323-483,na-sci-316-726}, which results in the circuit shown in Fig. 2d. The PPBSs have a different reflectivity $R$ and transmissivity $T$ for horizontal $H$ and vertical $V$ polarizations. We used three kinds of PPBSs: PPBS1 ($R_H = 50\%$, $R_V = 100\%$), PPBS2 ($R_H = 23\%, R_V = 100\%$), and PPBS3 ($T_H = 76\%, T_V = 100\%$). 
The control $C$ and target $T$ photons are first incident on PPBS1, which corresponds to BS5 in Fig. 2b, and two-photon quantum interference between the $H$ components occurs. The outputs are then routed to PPBS2 where quantum interference of the $H$ components with two auxiliary horizontally polarized photons occurs, which corresponds to BS7 and BS8 in Fig. 2b. The photons return to PPBS1 and a final quantum interference occurs, which corresponds to BS6 in Fig. 2b. The PPBS3 at each of the outputs implements 
BS9 and BS10 of Fig. 2b. The output of the CNOT is then detected by the photon counters with polarization analyzers. Note that all the four polarization modes of the control and target photons pass through all the optical components inside the interferometer so that the path difference between those four polarization modes are robust to drifts or vibrations of these optical components.

We used four photons generated via type-I spontaneous parametric down-conversion. The pump laser pulses (76 MHz at 390 nm, 200mW) pass through a beta-barium borate crystal (1.5 mm) twice to generate two pairs of photons. One pair was used as the $C$ and $T$ qubits, and the other as the auxiliary photons $A1$ and $A2$.
We first checked the quality of quantum interference \cite{ho-prl-59-2044} between a $C/T$ photon and an auxiliary photon at PPBS2. For example, to test the interference between $C$ and $A1$,  we detected photons $T$ and $A2$ just after the photon source to herald photons $C$ and $A1$, respectively, and measured the simultaneous single photon detection counts between detectors $DC$ and $DA1$ while scanning the arrival time of the $C$ photon. 
Note that the reflectivity of PPBS2 for horizontal polarization is 23\% and thus the visibility for perfect interference is $V_{th}=54\%$, rather than 100\% in the case of a 50\% reflectivity BS. The visibility $V_{exp}$ of the observed dips are $48\pm4\%$ and $49\pm3\%$ (with bandpass filters of center wavelength 780nm and FWHM 2nm), corresponding to relative visibilities of $V_{r}\equiv V_{exp}/V_{th}=$  $89\%$ and $91\%$. To test the performance of our CNOT gate circuit, we used coincidence measurements between the four threshold detectors at $DA1$, $DA2$, $DC$ and $DT$ rather than using photon number discriminating detectors for $DA1$ and $DA2$ because we needed to analyze the polarization state of the output to confirm correct operation. 
We performed this polarization analysis using a half-wave plate (HWP) or quarter-wave plate (QWP) together with a polarizing beam splitter (PBS).

\begin{figure}[t]
\begin{center}
\includegraphics*[width=8.5cm]{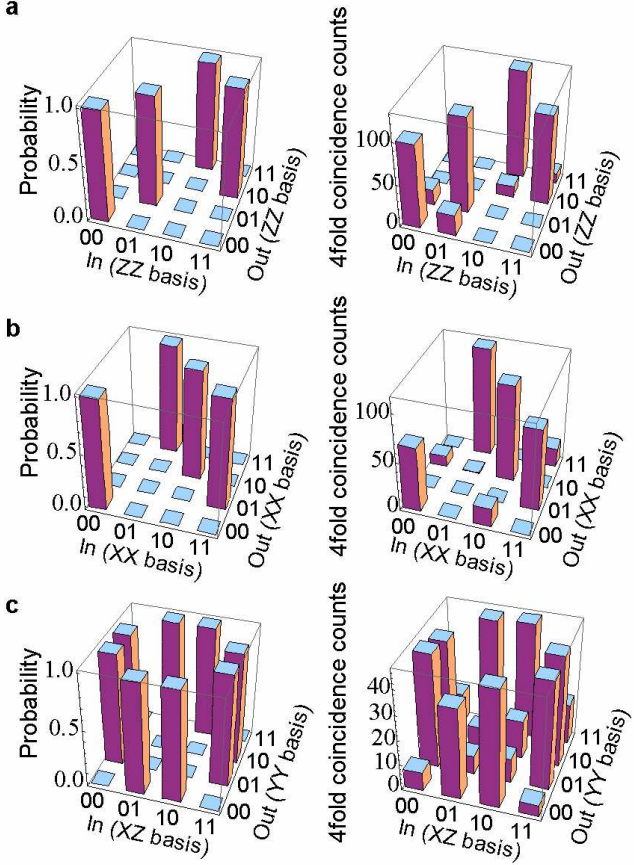}
\vspace{-0.3cm}
\caption{Experimental demonstration of a KLM CNOT gate. Left: ideal operation. Right: fourfold coincidence count rates (per 5000 s) detected at $DC$, $DT$, $DA1$ and $DA2$. 
(a) For control qubit, $|0_Z\rangle = |V\rangle$, $|1_Z\rangle = |H\rangle$; for target qubit, $|0_Z\rangle = 1/\sqrt{2}(|H\rangle + |V\rangle)$, $|1_Z\rangle = 1/\sqrt{2}(|H\rangle - |V\rangle)$. `10' indicates $C=1$ and $T=0$.
(b) For control qubit, $|0_X\rangle = 1/\sqrt{2}(|H\rangle + |V\rangle)$, $|1_X\rangle = 1/\sqrt{2}(|H\rangle - |V\rangle)$; for target qubit, $|0_X\rangle = |V\rangle$, $|1_X\rangle = |H\rangle$. 
(c) For control qubit, $|0_Y\rangle = 1/\sqrt{2}(|H\rangle + i|V\rangle)$, $|1_Y\rangle = 1/\sqrt{2}(|H\rangle - i|V\rangle)$; for target qubit, $|0_Y\rangle = 1/\sqrt{2}(|H\rangle + i|V\rangle)$, $|1_Y\rangle = 1/\sqrt{2}(|H\rangle - i|V\rangle)$.
 The events in which two pairs of photons are simultaneously incident to the ancillary inputs and no photons are incident to
the signal inputs are subtracted, as confirmed by a reference experiment without input photons}
\label{results}
\end{center}
\vspace{-0.8cm}
\end{figure}

We first checked the `logical basis' operation of the CNOT gate by preparing $C$ and $T$ in the four combinations of $|0\rangle$ and $|1\rangle$ (the $ZZ$ basis states) and measured the probability of detecting these $ZZ$ states in the output for each input state, to generate the `truth table' shown in Fig.~\ref{results}a.  The experimental data show the expected CNOT operation, \emph{i.e.} the $T$ photon's state is flipped only when the $C$ qubit is 1. The (classical) fidelity of this process $F_{ZZ \rightarrow ZZ}$, defined as the ratio of transmitted photon pairs in the correct output state to the total number of transmitted photon pairs, is $0.87\pm0.01$.
  
Next, we checked the average gate fidelity of our gate \cite{gi-pra-71-062310}. 
A measure of how all other possible gate operations (input and output states) perform is given by the average gate fidelity $\overline{F}$, which is defined as the fidelity of the output state averaged over all possible input states. This measure of the gate performance is given by \cite{gi-pra-71-062310}
\begin{equation}
  \overline{F}=(d~F_p+1)/(d+1)
\end{equation}
, where $d$ is the dimension of the Hilbert space ($d=4$ for a $2$ qubit gate). 
Because almost all the errors conserve horizontal/vertical polarization, the process fidelity $F_{P}$ is given by (see supporting online material)
\begin{equation}
F_{P}=(F_{ZZ \rightarrow ZZ}+F_{XX \rightarrow XX}+F_{XZ \rightarrow YY}-1)/2.
\end{equation}
Therefore, we need to obtain the fidelities of $F_{XX \rightarrow XX}$ and $F_{XZ \rightarrow YY}$ as well as $F_{ZZ \rightarrow ZZ}$. The measurement result of the input-output probabilities in the XX basis are shown in Fig.~\ref{results}b, where the basis states are \{$|0_X\rangle \equiv 1/\sqrt{2}(|0\rangle+|1\rangle)$, $|1_X\rangle \equiv 1/\sqrt{2}(|0\rangle-|1\rangle)$\}; the fidelity is $F_{XX \rightarrow XX}=0.88\pm0.02$. We also obtained $F_{XZ \rightarrow YY}$ from the experimental results shown in Fig.~\ref{results}c. The Y basis states are \{$|0_Y\rangle \equiv 1/\sqrt{2}(|0\rangle + i|1\rangle)$, $|1_Y\rangle \equiv 1/\sqrt{2}(|0\rangle + i|1\rangle)$\}. The fidelity is $F_{XZ \rightarrow YY}=0.81\pm0.02$. Based on eq.(1) and (2), our results show that the average gate fidelity of our experimental quantum CNOT gate is $\overline{F}=0.82\pm0.01$.


The data presented above confirm the realization of the CNOT gate proposed by KLM, which is an optical circuit combining a pair of efficient nonlinear elements induced by measurement. 
This confirms the first step in the KLM `recipe' for all-optical quantum computation and illustrates how efficient nonlinearities induced by measurement can be utilized for quantum information science; such measurement-induced optical nonlinearities could also be an alternative to nonlinear media required in a broader range of science.
For the present tests of the performance of CNOT gate operation, we used threshold detectors to monitor the output state. For applications in which the output state cannot be monitored, high-efficiency number-resolving photon detectors \cite{ki-apl-74-902} could be used at $DA1$ and $DA2$ to generate the heralding signals. 
Our device will be useful for conventional and cluster state approaches to quantum computing \cite{ni-prl-93-040503}, as well as quantum communication \cite{gi-nphot-1-165} and optical quantum metrology \cite{na-sci-316-726}. It could be implemented using an integrated waveguide architecture \cite{po-sci-320-646}, in which case a dual-rail encoding could conveniently be used.
\\

\noindent Acknowledgements:
We thank T. Nagata and M. Tanida for help and discussions. This work was supported by the Japan Science and Technology Agency (JST), Ministry of Internal Affairs and Communication (MIC), Japan Society for the Promotion of Science (JSPS), 21st Century COE Program, Special Coordination Funds for Promoting Science and Technology, EPSRC, QIP IRC, IARPA, ERC, and the Leverhulme Trust. J.L.O'B. acknowledges a Royal Society Wolfson Merit Award.


\clearpage

\section*{Appendix 1}

\textit{\section*{Derivation of the process fidelity}}

The PPBSs used to realize the KLM CNOT gate preserve the horizontal/vertical
polarization with high fidelity. In the quantum CNOT operation, these polarizations
correspond to the $ZX$-basis of the qubits. In the data shown in Fig. 3, this means
that the number of flips observed for the control qubit in 3A and for the target qubit in
3B are negligibly small. We can therefore describe the errors of the quantum gate
in terms of dephasing between the $ZX$-eigenstates. In terms of the operator expansion
of errors, we can define the correct operation $\hat{U}_{\mathrm{gate}}$ 
and three possible phase flip errors as
\footnotesize{
\begin{eqnarray}
\hat{U}_{\mathrm{gate}}=\ket{VV}\bra{VV}+\ket{VH}\bra{VH}+\ket{HV}\bra{HV}-\ket{HH}\bra{HH},
\nonumber \\
\hat{U}_{T}=\ket{VV}\bra{VV}-\ket{VH}\bra{VH}+\ket{HV}\bra{HV}+\ket{HH}\bra{HH},
\nonumber \\
\hat{U}_{C}=\ket{VV}\bra{VV}+\ket{VH}\bra{VH}-\ket{HV}\bra{HV}+\ket{HH}\bra{HH},
\nonumber \\
\hat{U}_{CT}=\ket{VV}\bra{VV}-\ket{VH}\bra{VH}-\ket{HV}\bra{HV}-\ket{HH}\bra{HH}.
\end{eqnarray}
}
The operation of the gate can then be written as 
\begin{equation}
\label{eq:pm}
E(\rho_{in})= \sum_{n,m} \chi_{nm} \hat{U}_n \rho_{in} \hat{U}_m
\end{equation}
where $n,m \in \{\mathrm{gate}, T, C, CT\}$ and $\chi_{nm}$ define the process matrix
of the noisy quantum process.

Each of our experimentally observed truth table operations $i\to j$ is 
correctly performed by
$\hat{U}_{\mathrm{gate}}$ and one other operation $\hat{U}_n$.
Therefore, the fidelities $F_{i\to j}$ can be given by the sums of the probability
$F_p=\chi_{\mathrm{gate},\mathrm{gate}}$  for the correct operation $\hat{U}_{\mathrm{gate}}$ and the probabilities
$\eta_{n} = \chi_{nn}$ for the errors $\hat{U}_n$ as follows.
\begin{eqnarray}
F_{ZZ \to ZZ} &=& F_p+\eta_{T}
\nonumber
\\
F_{XX \to XX} &=& F_p+\eta_{C}
\nonumber
\\
F_{XZ \to YY} &=& F_p+\eta_{CT}
\label{eq:fid}
\end{eqnarray}
Note that these relations between the diagonal elements of the process 
matrix and the experimentally observed fidelities can also be derived from
eq. (\ref{eq:pm}) using the formal definition of the experimental fidelities. 
In this case the fidelities are determined by the sums over the correct outcomes
$|(j)_l \rangle$ in $E(|(i)_k\rangle \langle(i)_k|)$, averaged over all inputs $|(i)_k \rangle$,
\begin{eqnarray}
F_{i\to j} &=& \sum_{l,k} \langle(j)_l| E( |(i)_k\rangle \langle(i)_k| ) |(j)_l \rangle/4) 
\nonumber
\\
&=& \sum_{n,m} \chi_{nm} (\sum_{l,k} \langle(j)_l|\hat{U}^\dagger_n|(i)_k \rangle \langle(i)_k|\hat{U}_m|(j)_l \rangle/4).
\end{eqnarray}
Here $k,l \in \{1,2,3,4\}$, and $(i)_k$ denotes the $k$ th state of the $i$ basis states. For example, $|(i)_1 \rangle=|VV \rangle, |(i)_2 \rangle=|VH \rangle, |(i)_3 \rangle=|HV \rangle, |(i)_4 \rangle=|HH \rangle$ for $i=ZX$.
The sums over initial states $k$
and final states $l$ are one for $n=m=0$ and for a single other error, $n=m=n(ij)$.
All remaining sums are zero, confirming the results in eq.(\ref{eq:fid}).

Since the diagonal elements of the process matrix correspond to the
probabilities of the orthogonal basis operations, their sum is normalized to one,
so that $\sum_n \chi_{nn}=F_p+\eta_{T}+\eta_{C}+\eta_{CT}=1$. 
It follows that the sum of all three experimentally determined fidelities is
$F_{ZZ \to ZZ}+F_{XX \to XX} + F_{XZ \to YY} = 2 F_p+1$. Therefore, the process 
fidelity of our KLM CNOT gate is given by
\begin{equation}
F_p=(F_{ZZ \to ZZ}+F_{XX \to XX}+F_{XZ \to YY}-1)/2 = 0.78.
\end{equation}
This clearly exceeds the threshold $F_p\ge0.5$ for the gate to produce entanglement --- a key quantum operation of the gate. 
The fidelity of the output states of the gate, averaged over all input states is related to the process fidelity 
\begin{equation}
\overline{F}=(dF_p+1)/(d+1)=0.82
\end{equation}
where $d$ is the dimension of the Hilbert space (d=4 for a two qubit gate).

\end{document}